# PMUT-Powered Photoacoustic Detection: Revolutionizing Microfluidic Concentration Measurements

Kaustav Roy[1], Akshay Kumar[2], Vijayendra Shastri[3], Isha Munjal[4], Kritank Kalyan[5], Anuj Ashok[6], Anthony Jeyaseelan[7], Jaya Prakash[4†], and Rudra Pratap[2†]

**Abstract—** This report introduces a novel optofluidic platform based on piezo-MEMS technology, capable of identifying subtle variations in the fluid concentration. The system utilizes piezoelectric micromachined ultrasound transducers (PMUTs) as receivers to capture sound waves produced by nanosecond photoacoustic (PA) pulses emanating from a fluid target housed in PDMS microchannels. Additionally, a dedicated low-noise single-channel amplifier has been developed to extract the minute analog voltage signals from the PMUTs, given the inherently weak ultrasound signals generated by fluid targets. The PMUTs' proficiency in detecting changes in fluid concentration under both static and time-varying conditions has been documented and verified, confirming the platform's efficacy in monitoring fluid concentrations.

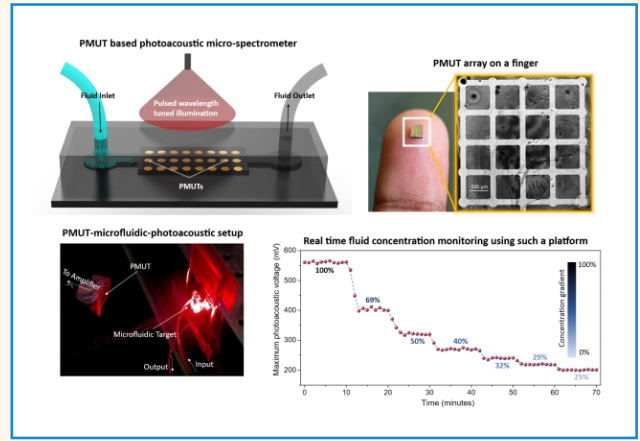

This work was funded from the grant – STARS/APR2019/NS/653/FS from the MOE.

K. R. was with the Centre for Nano Science and Engineering, Indian Institute of Science, Bangalore, 560012, India (e-mail: kaustav@alum.iisc.ac.in).

A. K. is with the Centre for Nano Science and Engineering, Indian Institute of Science, Bangalore, 560012, India (e-mail: akshaykumar1@iisc.ac.in).

V. S. is with the Department of Precision and Microsystems Engineering, TU Delft, 2628 CD Delft, Netherlands (e-mail: V.U.Shastri@tudelft.nl).

I. M. is with the Department of Instrumentation and Applied Physics, Indian Institute of Science, Bangalore, 560012, India (e-mail: isha17774@iisc.ac.in).

K. K. is with the Singh Center for Nanotechnology, University of Pennsylvania, PA 19104, US (e-mail: kritank@upenn.edu).

A. A. is with the Elmore Family School of Electrical and Computer Engineering, Purdue University, IN 47907, US (e-mail: anujashok@gmail.com).

A. J. is with the Centre for Nano Science and Engineering, Indian Institute of Science, Bangalore, 560012, India (e-mail: antonya@iisc.ac.in).

J. P. is with the Department of Instrumentation and Applied Physics, Indian Institute of Science, Bangalore, 560012, India (e-mail: jayap@iisc.ac.in).

R. P. is with the Centre for Nano Science and Engineering, Indian Institute of Science, Bangalore, 560012, India (e-mail: pratap@iisc.ac.in).

†: Corresponding author

## I. INTRODUCTION

LIQUIDS are prevalent in numerous sectors including pharmaceuticals, oil and gas, petrochemicals, automotive, and various processing industries, serving diverse practical needs. The dye industry, a significant player, produces and utilizes liquids in the creation of dyes for multiple applications like food, textiles, photography, and leather. This industry relies on solvents to adjust dye concentrations. However, it's known that the concentration of the colorant in a dye-solvent combination diminishes over time [1]. Consequently, monitoring dye concentration is crucial, yet there are only a handful of studies addressing this issue [2].

The realm of transducers has experienced multiple transformative phases, one of which was sparked by the advent of nanotechnology. This innovation facilitated the production of ultra-small transducers having submicron dimensions. This breakthrough, coupled with advanced material science, led to the emergence of microelectromechanical systems (MEMS) [3]. This transition marked a shift from the macroscopic sensor world to a microscale one, prompting a surge in demand within the field. Nowadays, MEMS inertial systems are a common feature in the majority of smartphones [4]. Furthermore, MEMS has had a groundbreaking impact on ultrasound technology, culminating in the development of exceedingly small ultrasound transducers. Commonly referred to as micromachined ultrasound transducers (MUTs), these devices

*Highlights*

- **A PMUT based optofluidic platform is presented which employs photoacoustics to detect micro changes in target fluid concentration**

- **The system was observed to bear a sensitivity of 4.8 mV/% change in concentration from an effective cross-section of 0.05 mm²**

- **This research lays the foundation to build futuristic all-integrated on-chip photoacoustic spectrophotometers for fluid characterization and detection** come in two primary types: capacitive MUTs (CMUTs) [5]–[8] and piezoelectric MUTs (PMUTs) [9]–[23]. Furthermore, nanotechnology has also facilitated the development of miniature fluidic devices, such as microvalves [24], micropumps [25], microfluidic flow sensors [26], spaces to generate sound [34]–[37]. This process, known as photoacoustics, involves producing sound from an object exposed to pulsating light and finds numerous applications across various sectors [38]–[47]. The combination of light's electromagnetic characteristics and sound's penetration

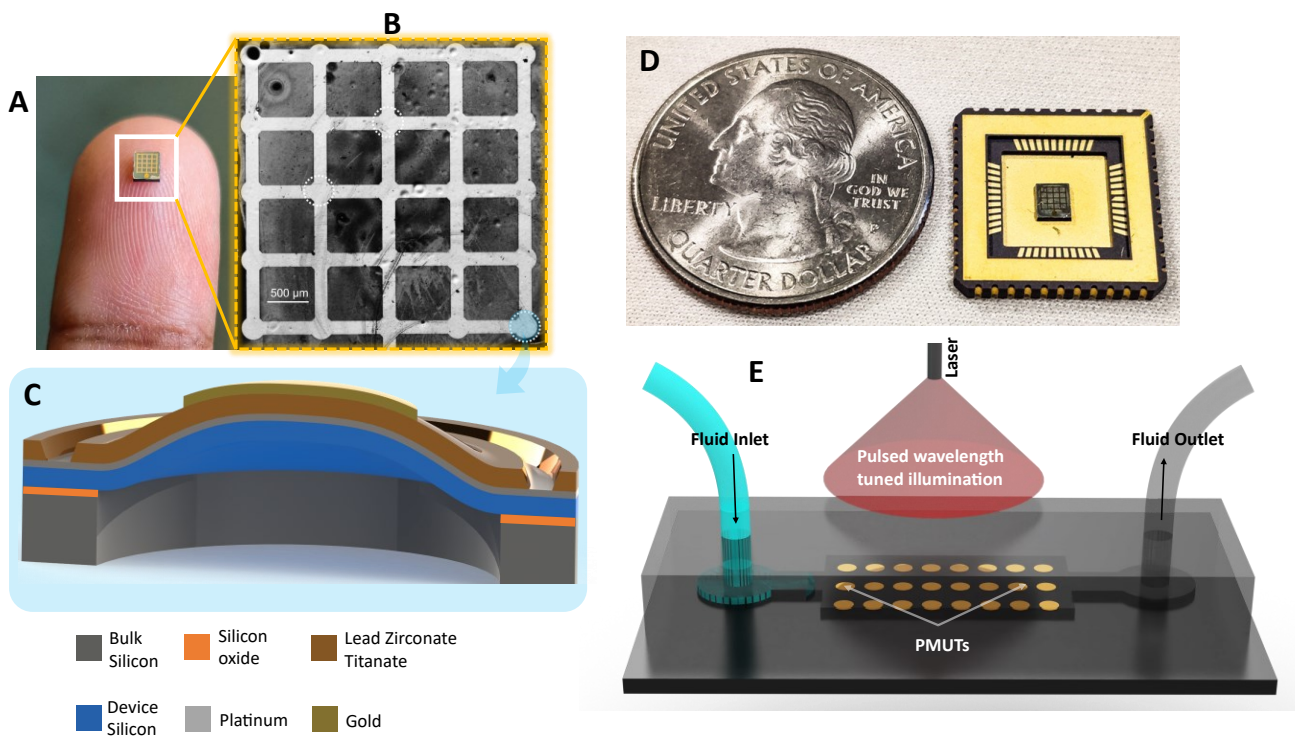

Figure 1. PMUT based photoacoustic detection. A. Picture of the PMUT die used in the work. B. A zoomed in microscopic view of the PMUT die containing 25 cells connected together to form one single element. C. Stacked 2D schematic showing the constitutive layers of a PMUT. D. The PMUT die packaged in a chip carrier. E. Concept futuristic schematic of PMUT integrated microfluidic device with an external tunable pulsed light source as a photoacoustic spectrophotometer.

microneedles [27], and micronozzles [28], among others.

The integration of light with microfluidic systems has given birth to an innovative field known as optofluidics [29]. This domain leverages the benefits of fluidics — like controlled mixing of soluble liquids, the formation of stable interfaces between immiscible fluids, and excellent medium for transport — along with the strengths of optics, such as highly precise optical sensing methods, the capability to focus light at sub-micron levels, and object manipulation. This synergy has spurred the development of cutting-edge devices, including fluid lenses [30], optofluidic microscopes [31], resonators [32], and tweezers [33], among others. The introduction of pulsed or modulated light sources has further expanded this field, influencing the fluids contained in micro capacity allows for deeper exploration and gathering of valuable data from the targeted subjects.

Recently, there has been a surge of contributions in making PMUT based photoacoustic devices primarily for deep tissue imaging applications [48]–[54]. The utility of using PMUTs stems from their several advantages over the traditional bulk piezo based transducers such as – tiny, scalable form factor, minimal power requirement, less inter-device variability, CMOS compatibility, manufacturability in bulk batches. They are fabricated by using precise nanofabrication tools which gives them a sheer benefit over the ubiquitous bulk ceramic-based ultrasound transducers, since the accuracy of such tools to create the tiniest design implementation is unparallel.

There has been a few works with PMUTs applied to concentration sensing particularly to detect sugar concentration in solution [55], however, such a scheme uses an ultrasound only approach unlike the photoacoustic modality as used in this work. Light can probe the fluid of interest electromagnetically and can bring in more subtle information in terms of concentration levels thereby assisting in the detection of concentration more effectively.

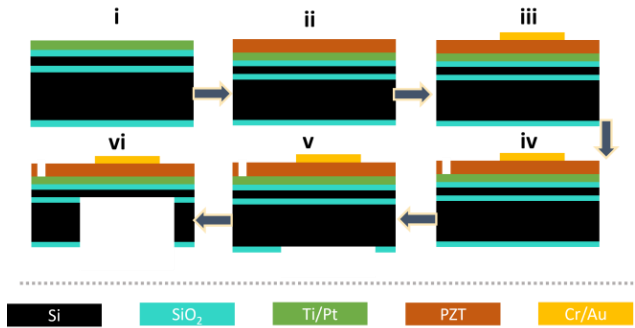

Figure 2. Fabrication process flow for a PZT based thin film PMUT.

| | | | | |
|---|---|---|---|---|
| Si | SiO₂ | Ti/Pt | PZT | Cr/Au |

In this work, we developed the functional thin-film piezoelectric material such as the lead zirconate titanate and employed nanofabrication techniques to create PMUTs which were then used as photoacoustic receivers. Additionally, we constructed microfluidic channels from PDMS-glass, serving as flexible vessels for fluid targets. We then established an experimental framework that incorporated a pulsed laser, a microfluidic channel, and PMUTs to validate their potential in receiving photoacoustic signals. Custom low noise amplification circuit was also developed which formed a crucial component in assisting with the smooth retrieval of the weak PA signals from the PMUT receiver. Upon evaluation, PMUTs proved to be viable for photoacoustic endeavours. We proceeded with experiments using various concentrations of commercially available blue ink mixed with deionized (DI) water to serve as fluid photoacoustic targets, confirming the PMUT's ability to discern concentration changes as the ink levels in the DI water were altered. We then extended our effort to build a time dynamic concentration monitoring system which could demonstrate the real time mixing dynamics, thereby demonstrating the successful application of PMUTs as detectors of species concentration.

## II. PMUT AND MICROFLUIDIC CHANNELS USED IN THE WORK

### A. PMUTs

Piezoelectric Micromachined Ultrasound Transducers (PMUTs) are acoustic devices driven by thin film piezoelectricity and are created by using micro/nanofabrication techniques. Essentially, PMUTs are composed of microplates (vibration mechanics governed by the flexural rigidity) or micro-membranes (vibration mechanics governed by tension) with different distinct layers arranged atop one another, backed by an acoustic cavity. The PMUTs used in this work for conducting the photoacoustic studies were developed at the National Nanofabrication Centre (NNFC), part of the Centre for Nano Science and Engineering (CeNSE) at the Indian Institute of Science (IISc) in Bangalore. Figure 1A depicts the PMUT die

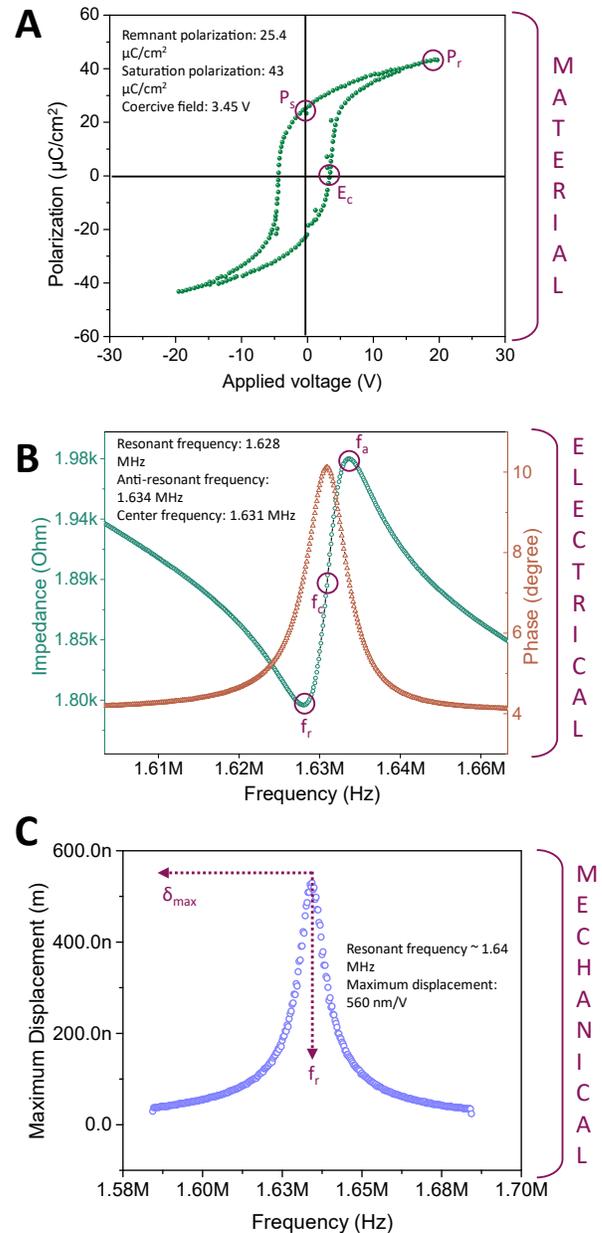

Figure 3. PMUT characterization. A. Material: polarization vs. voltage hysteresis loop, B. Electrical: absolute electrical impedance and phase vs. frequency, and C. Mechanical: maximum displacement of the PMUT vs. frequency.

used in this work while being held on a finger and Figure 1B depicts a zoomed in microscopic picture of the die showing a connected matrix of 25 PMUT cells which work together to form an element. The PMUT consists of six different thin film layers as depicted in Figure 1C, such as the bulk silicon, silicon-di-oxide, device silicon, platinum, lead zirconate titanate and gold. The bulk silicon serves as the handle layer, whereas the device silicon serves as the structural layer, thereby giving the PMUT most of their structural properties. The thin film of lead zirconate titanate (PZT) acts as the active layer which either vibrates the diaphragm or gets charged upon diaphragm movement. Consequently, a PMUT can function as a transmitter, receiver, or both. The PZT thin film was deposited onto a silicon-on-insulator

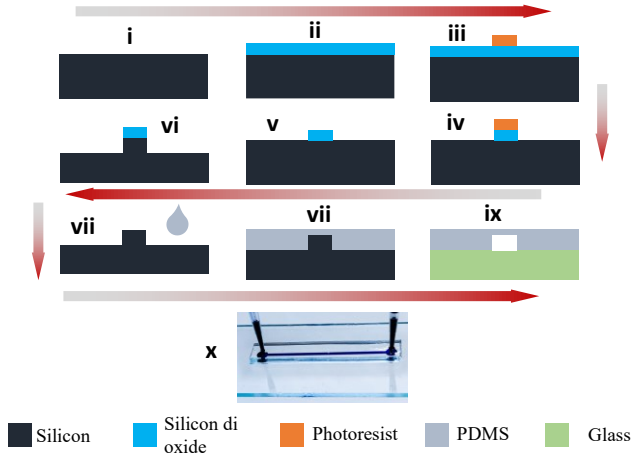

Figure 4. Process flow for making the microfluidic channel on glass

**Legend:** ■ Silicon  ■ Silicon di oxide  ■ Photoresist  ■ PDMS  ■ Glass

(SOI) substrate with a uniform device layer thickness of 10 μm. A PMUTs' properties can be customized by adjusting the device layer's dimensions, modifying the combined layer thickness, and altering the fabrication stresses between layers. Post-fabrication, the PMUT was connected to a ceramic package with a pin grid array and covered with parylene-c to improve resistance to noise, as illustrated in Figure 1D. A futuristic representation of the microfluidic integrated micro photoacoustic device having an array of PMUTs integrated as receivers is depicted in Figure 1E. With a wavelength tunable light source, such a device can be used as an on-chip spectrophotometer to detect and analyze fluid samples.

### B. PMUT Fabrication

The PMUTs used in this study were produced following the procedure depicted in Figure 2. This method begins with a silicon-on-insulator wafer coated with titanium/platinum and featuring a 10 μm thick device layer (Figure 2i). Through multiple repetitions of the sol-gel method, lead zirconate titanate (PZT) is spin-coated to attain the required thickness (approximately 0.5 μm), followed by an annealing process at 650°C to create a consistent thin film (Figure 2ii). Next, the upper electrode (Chrome/Gold – 30/120 nm) is patterned using lithography (Figure 2iii), and the PZT layer is subsequently removed through wet etching (Figure 2iv). The composite is then patterned from the rear to eliminate the backside oxide using reactive ion etching (RIE), as illustrated in Figure 2v. In the final steps, the supporting layer is etched via deep reactive ion etching (DRIE) to release the devices, and the buried oxide is etched away to relieve any residual stress in the released stack Figure 2vi.

### C. PMUT Characterization

Post piezoelectric poling, one of the typical cells in the PMUT die is characterized materially, to understand the properties of the piezoelectric thin film, electrically, to determine the electrical impedance of the device and mechanically, to determine the device deflection sensitivity. The electrical quality of the PZT thin film was evaluated through polarization, achieved by varying the voltage using a Precision Materials Analyzer from Radiant Technologies Inc.

The resulting hysteresis loop is depicted in Figure 3A, displaying the characteristic ferroelectric switching of the PZT thin film in response to the applied electric field. Key parameters extracted from the graph include the remanent polarization (Pr), saturated polarization (Ps), and the coercive field (Ec), measured at 25.4 μC/cm², 43 μC/cm², and 3.45 V, respectively. To understand the AC characteristics of the device, electrical characterization was conducted using 4294A Precision Impedance Analyzer, Agilent Inc., and the results obtained are shown in Figure 3B. The frequency was swept from 1.61 MHz to 1.66 MHz. The maximum impedance magnitude was found to be 1.98 kΩ and the minimum impedance magnitude was found to be 1.80 kΩ respectively. The resonant frequency ($f_r$), center frequency ($f_c$) and the antiresonant frequency ($f_a$) was found to be 1.628 MHz, 1.631 MHz, and 1.634 MHz respectively. The effective electromechanical coupling coefficient has been determined to be 0.73%. From the phase spectra, a change in 6 degrees has

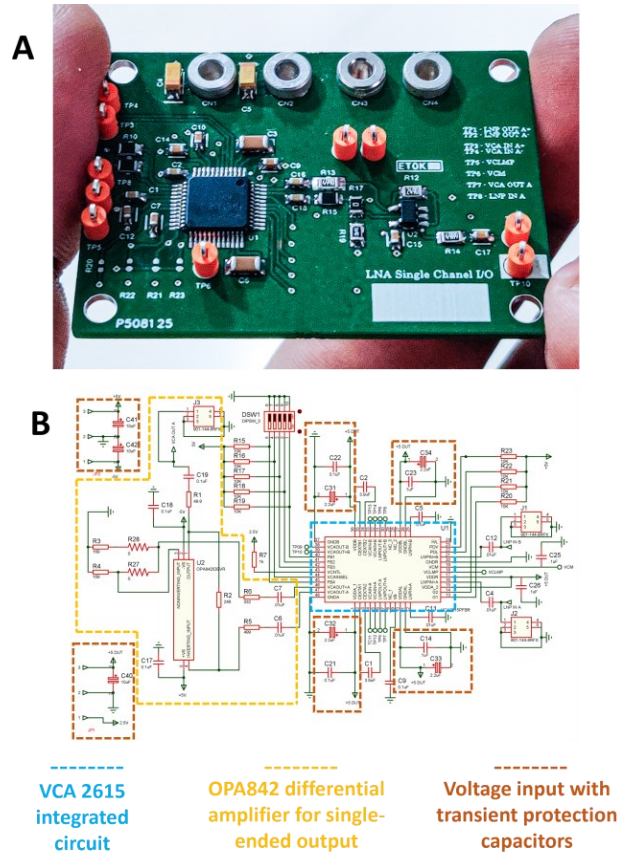

**VCA 2615 integrated circuit** · **OPA842 differential amplifier for single-ended output** · **Voltage input with transient protection capacitors** been observed at resonance. To understand the behaviour of the PMUTs mechanically, one of the typical PMUTs were selected from the die and was characterized using the laser doppler vibrometer (Microsystems Analyzer, MSA 500) from Polytec Inc (Figure 3C). The PMUT cell demonstrated a maximum peak hold deflection of 560 nm/V at a frequency of 1.64 MHz.

Figure 5. PMUT signal amplification circuit. A. Printed circuit board containing the VCA 2615 IC from the Texas Instruments, along with the additional support circuitry. B. Schematic of the circuit showing connections in detail.

### D. Fabrication of Microfluidic Devices

The microfluidic channel was developed at NNFC, CeNSE using a series of steps outlined below. Initially, a silicon mold underwent RCA cleaning (Figure 4i), followed by the addition of silicon dioxide to act as a robust mask for later etching

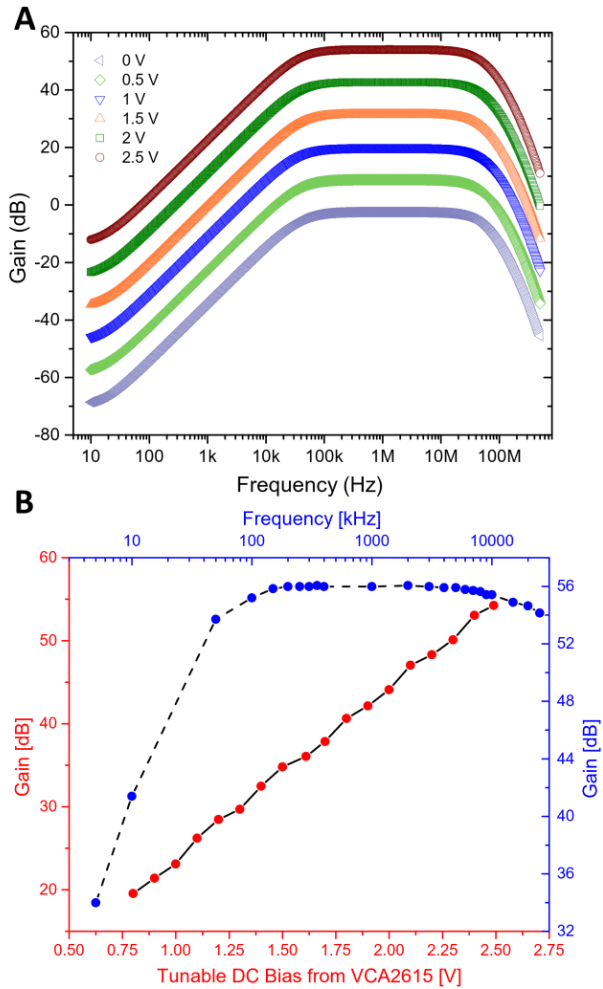

**A**

**B**

Figure 6. PMUT amplifier in simulation and characterization. A. LT Spice simulation of the designed circuit with a broad gain band from 100 kHz to 10 MHz. B. Experimental validation by characterizing the amplifier PCB.

phases (Figure 4ii). The substrate that resulted was then coated with a thick layer of AZ4562 photoresist (Figure 4iii), after which oxide etching was performed in areas without patterning (Figure 4iv) using reactive ion etching. Subsequently, the photoresist was removed through ashing (Figure 4v). The substrate then underwent a timed deep reactive ion etching process to achieve a specific depth (Figure 4vi), creating the microfluidic mold. Afterward, PDMS, combined with a curing agent in a 10:1 ratio, was poured over the mold (Figure 4vii) and set at 120 °C for 30 minutes. The solidified PDMS was gently peeled off the mold (Figure 4viii) and adhered to a glass slide with a plasma etcher from Harrick Plasma Inc (Figure 4ix). The assembled microfluidic apparatus was then heated at 120 °C to reinforce the bond. Figure 4x shows the standard bonded microfluidic device used in the experiment, featuring cross-sectional dimensions of 1000 μm in width and 50 μm in depth.

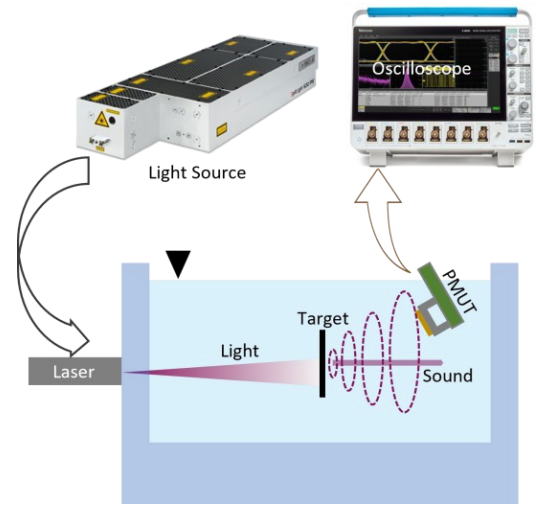

Figure 7. A schematic of PMUT-PA experimental setup

### III. PMUT PHOTOACOUSTIC INSTRUMENTATION

#### A. PMUT Signal Amplification

In house, low noise Amplifier (LNA) circuit was designed for low noise PMUT PA signal amplification as shown Figure 5A. Figure 5B shows the schematic of the circuit. The heart of this circuit consists of a Texas Instrument's VCA2615 IC [56] for excellent input signal handling capabilities, combining low-noise

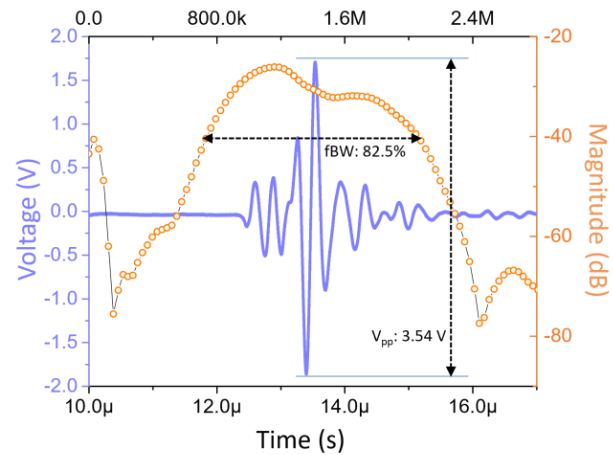

Figure 8. PA A-line time and frequency domain signal as obtained from the PMUT. The -6dB fractional bandwidth was observed to be 82.5%. characterizing the PCB

preamplifier (LNP) and a voltage-gain amplifier (VGA) inside the IC. The differential output of VCA 2615 is buffered using a separate OPA842 op-amp in differential amplifier configuration for single ended output. A +/-5V power supply is used for the LNA circuit and a DC bench top power supply is used to control the $V_{cntl}$, which ultimately controls the VGA gain with the linear control response of 22 dB/V. G1 and G2 are held high for 22 dB LNP gain settings as at this setting achieves 0.7 nV/√Hz voltage noise and typically 1 pA/√Hz current noise. H/L pin held low for +6 dB discrete gain. For limiting the amplified output voltage swing, $V_{clmp}$ can be used as input voltage reference for output clipping to desired voltage level. Figure 6 shows the characterization of the constructed amplifier circuit by simulation (see Figure 6A) and by actual experiment (see Figure 6B). The simulation and the experiment tests closely match with a broad gain band from 100 kHz to 10 MHz of frequencies. The linearity on changing the magnitude of the $V_{cntl}$ is also observed in both the simulation and the experiment with the gain rising from ~20 dB at ~0.75 V to ~55 dB at 2.5 V.

### B. The Photoacoustic Setup

Figure 7 illustrates a diagram outlining the setup for the photoacoustic experiment. This system includes a versatile OPO nanosecond pulsed laser (specifically, the SpitLight 1000 from InnoLas Lasers GmbH), adjustable within the 660 nm to 2500 nm wavelength range. This laser has a 7 ns pulse low-noise, voltage-controlled amplifier (VCA 2615, Texas Inc.), built inhouse, before being transmitted to an oscilloscope (Mixed Signal Oscilloscope, Tektronix Inc.). This signal is then relayed to a computer for immediate, digitally conditioned visualization. Precise positional adjustments of the PMUT are made possible through an x-y linear stage (CONEX-TRA25CC, Newport Inc.), with movements directed by specialized software on the computer. Additionally, the computer governs the pulsed laser, producing precise triggers that are in sync with the oscilloscope, thereby facilitating exact dynamic assessments.

## IV. RESULTS

### A. The PA A-line

In this study, the packaged PMUT was evaluated as an acoustic sensor using the previously described experimental arrangement. A pencil lead with a circular cross-section and a diameter of 500 width and operates at a repitition rate of 30 Hz. The laser targets a photoacoustic target (microfluidic sample or pencil lead) affixed to the side of a specially designed glass tank, using a 660 nm wavelength. Upon absorbing the fluctuating light, the target generates acoustic signals, which the PMUT detects. The PMUT's voltage output is first amplified by a μm served as the target for photoacoustic emissions during laser exposure to pulsed nanosecond light, positioned 5 cm from the PMUT's surface. The collected data is presented in Figure 8. The PMUT registered a peak-to-peak voltage of 3.54 V on the oscilloscope after amplification by 55 dB, equating to an original signal of 6.29 mV from the PMUT. An A-line data analysis was conducted using a Fast Fourier Transform (FFT), with the results depicted in Figure 8. The predominant frequency in water was identified as 1.12 MHz. Additionally, the fractional -6 dB bandwidth was computed, revealing a value of 82.5%.

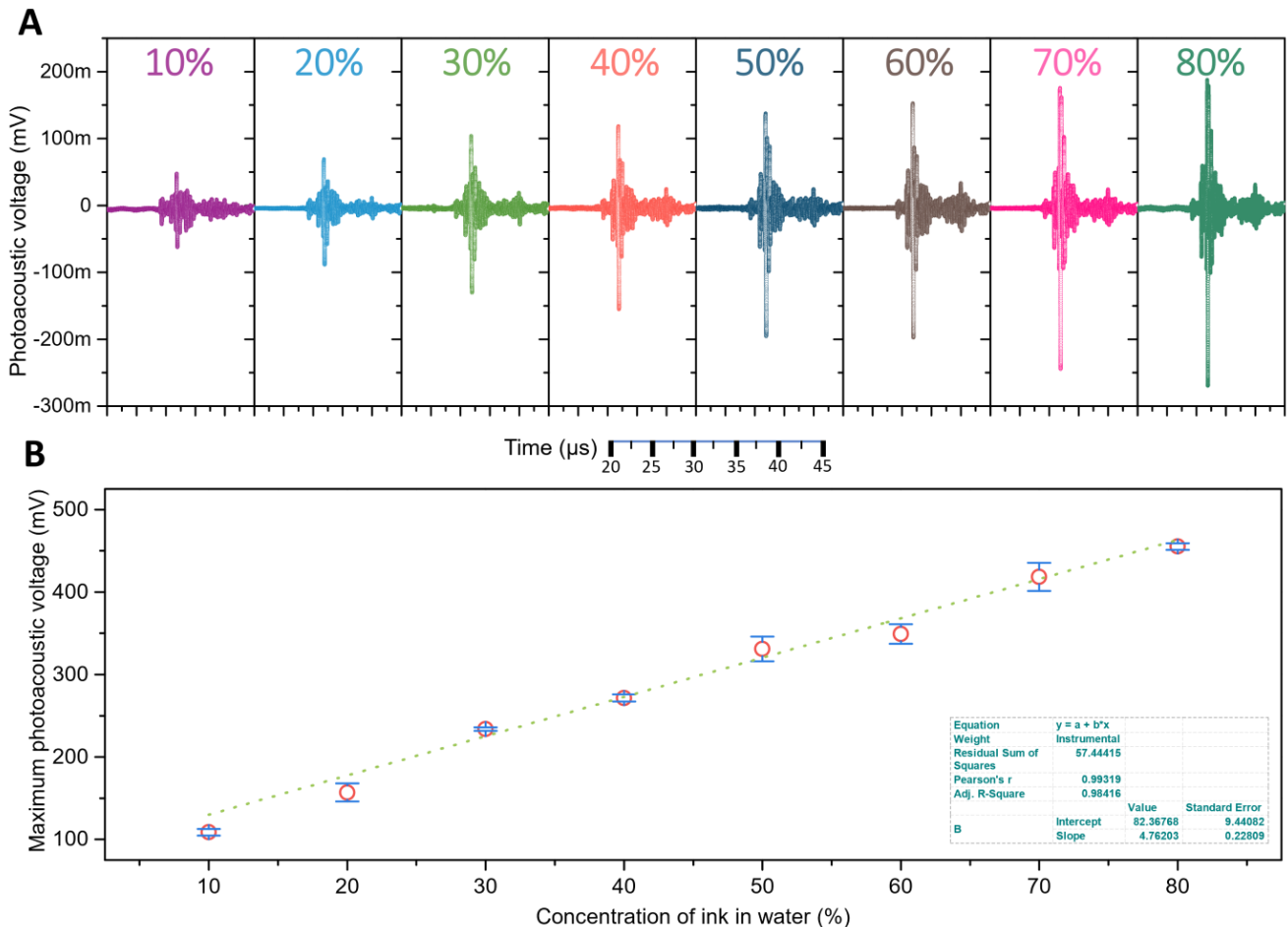

Figure 9. Static concentration measurements. A. PA A-line voltage signals as received from the PMUT for ink-water fluid targets confined in microfluidic channels. B. Maximum PA signal amplitude plotted with ink concentration.

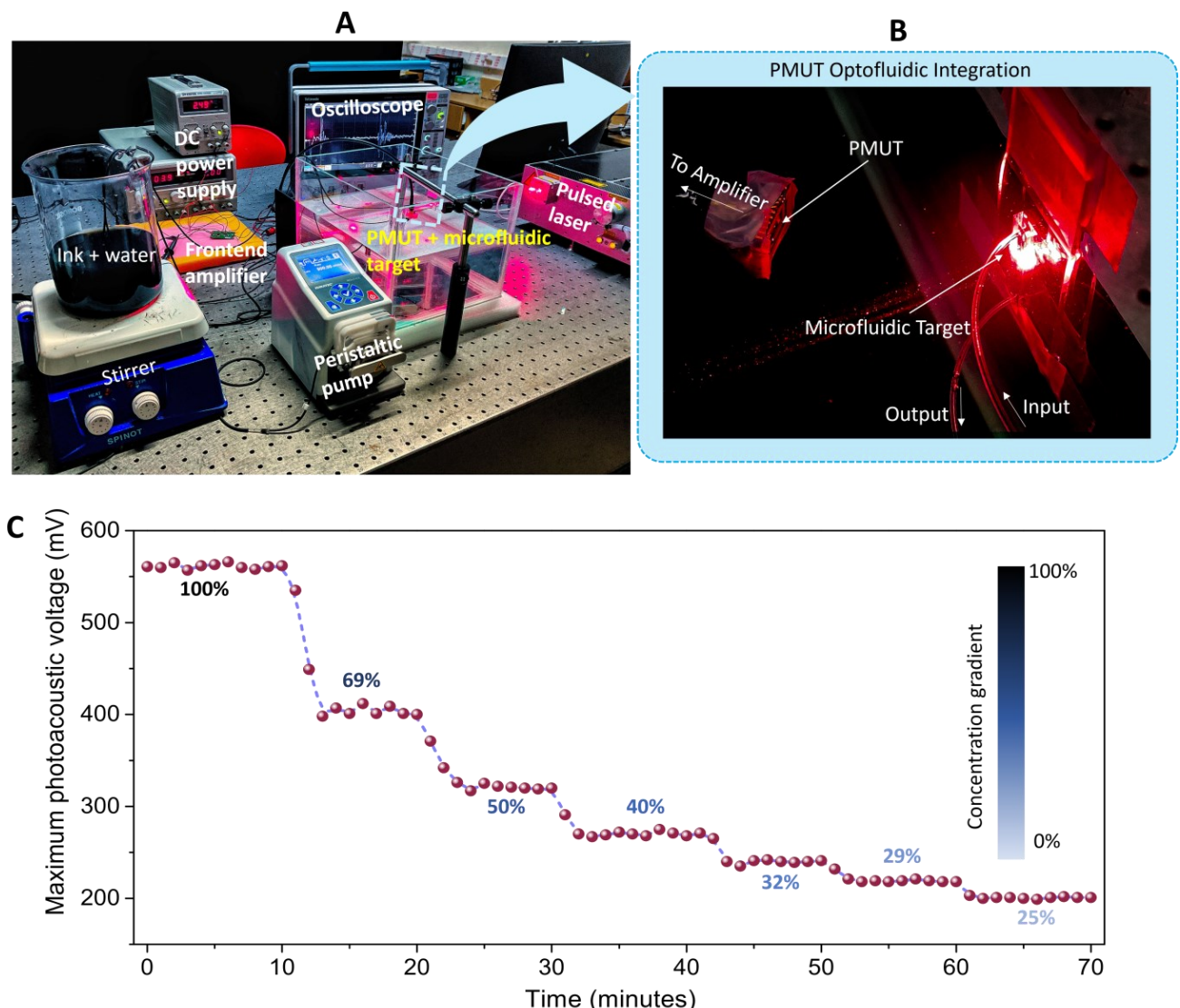

Figure 10. Real time concentration measurement. A. Experimental setup for conducting the real time concentration monitoring. B. A zoomed in view of the PMUT PA microfluidic system. C. Concentration time dynamics as observed using the PMUT, captured from the fluid inside the microfluidic channel.

### B. Static Concentration Measurements

The PMUT PA microfluidic concentration detector was next used to detect the change in concentration for particular fluid mixtures. Blue colored India ink along with DI water was used as the solution which was injected into the microfluidic target. Eight different concentrations were prepared from 10% to 80% in steps of 10%. The time domain PA A-line obtained from each target concentrations are depicted in Figure 9A. All of the A-line signals follow the same observable pattern. The peak-to-peak voltage magnitude was extracted from the A-lines obtained from each different concentration and plotted as depicted in Figure 9B. A linear fit with a r squared of 0.99 was obtained suggesting a linear relationship between the peak-to-peak voltage magnitude and the species concentration. The slope of the fitted line suggests a sensitivity of 4.8 mV/% change in concentration.

### C. Real-time Concentration Measurements

Next, in order to establish the PMUT PA-microfluidic platform as a real time concentration detector, the setup was enhanced by the addition of a mixture reservoir with an associated stirrer and a peristaltic pump Reglo ICC from Ismatec Inc. to circulate the reservoir fluid real time through the microfluidic channel, at a flow rate of 60 mL/min while illuminating the channel constantly with the pulsed light in the wavelength of 660 nm. Figure 10A depicts the experimental setup used for the real time concentration measurements showing the necessary parts. Figure 10B represents the zoomed-in view of the PMUT optofluidic integration showing the PMUT, light and the microfluidic target. The plot obtained from Figure 10C shows the time dynamics of the concentration change as tracked by the PMUT PA microfluidic detector. The experiment was initialized with 200 mL of a blue colored India ink in the reservoir which was circulated through the microfluidic channel for the first 10 minutes. The maximum PA voltage obtained was noted at ∼ 550 mV. After 10 minutes, 100 mL of DI water was added to the reservoir. The voltage was observed to drop for the next 4 minutes, thereby reaching saturation at ∼ 400 mV. The concentration of ink to water at this level of mixture was 69%. DI water was continually added after every 10 minutes till 70 minutes of time until the real time voltage to time curve hit saturation. As the experiment progressed, with each new addition in water the magnitude of the decrease in voltage was found to be lower as compared to the previous addition, revealing the nonlinear nature of the mixing.

## V. CONCLUSION

In summary, we present a novel optofluidic system rooted in piezo-MEMS technology, designed to detect minor changes in fluid concentration. The setup employs PMUTs to receive sound waves generated by nanosecond photoacoustic (PA) pulses from a fluid sample contained within PDMS microchannels. To aid in capturing the weak analog voltage signals from the PMUTs, due to the typically low-intensity ultrasound signals from fluid samples, a specialized low-noise single-channel amplifier was crafted. The capability of the PMUTs to sense fluid concentration alterations, both in steady-state and dynamic scenarios, has been thoroughly assessed and validated, underscoring the system's effectiveness in fluid concentration tracking.

## ACKNOWLEDGMENT

The authors are grateful to all the staff at the National Nanofabrication Centre (NNFC), Micro Nano Characterization Facility (MNCF) at the Centre for Nano Science and Engineering (CeNSE), Indian Institute of Science, the funding agencies – the Department of Science and Technology (DST NanoMission), the Ministry of Education (MOE), the Ministry of Electronics, Information Technology (MeitY) and DBT-IYDF award. This work was funded from the grant – STARS/APR2019/NS/653/FS from the MOE.